\newif\ifsinglecolumn
\singlecolumnfalse

\ifsinglecolumn
	\documentclass[journal,onecolumn,draftclsnofoot]{IEEEtran}
\else
	\documentclass[journal]{IEEEtran}
\fi

\usepackage[utf8]{inputenc}
\usepackage[T1]{fontenc}
\usepackage{graphicx}
\usepackage{float}
\usepackage{booktabs}
\usepackage{amsmath}
\usepackage{url}
\usepackage{cite}
\usepackage{orcidlink}
\hypersetup{hidelinks}

\newif\ifpreprint
\preprinttrue

\newcommand{\rseiLatCpuDevFpThirtyTwo}{31.73}
\newcommand{\rseiLatCiCpuDevFpThirtyTwo}{3.81}

\newcommand{\rseiLatCpuDevIntEight}{20.75}

\newcommand{\rseiSpeedupIntEightCpuDev}{1.53}

\newcommand{\rseiSinconDeltaPctDevCpuFpThirtyTwo}{$-$5.1}
\newcommand{\rseiSinconDeltaPctDevCpuFpSixteen}{$-$8.2}
\newcommand{\rseiSinconDeltaPctDevCpuIntEight}{+12.2}

\newcommand{\rseiLatCpuCoffeeFpThirtyTwo}{33.80}

\newcommand{\rseiEraplCpuCoffeeFpThirtyTwo}{1.260}
\newcommand{\rseiEplugCpuCoffeeFpThirtyTwo}{1.723}

\newcommand{\rseiLatCpuCoffeeFpSixteen}{35.74}

\newcommand{\rseiEraplCpuCoffeeFpSixteen}{1.250}
\newcommand{\rseiEplugCpuCoffeeFpSixteen}{1.830}

\newcommand{\rseiLatCpuCoffeeIntEight}{26.56}

\newcommand{\rseiEraplCpuCoffeeIntEight}{0.859}
\newcommand{\rseiEplugCpuCoffeeIntEight}{1.230}

\newcommand{\rseiEraplGpuCoffeeFpThirtyTwo}{0.865}
\newcommand{\rseiEplugGpuCoffeeFpThirtyTwo}{1.268}

\newcommand{\rseiEraplGpuCoffeeFpSixteen}{0.861}
\newcommand{\rseiEplugGpuCoffeeFpSixteen}{1.263}

\newcommand{\rseiEraplGpuCoffeeIntEight}{1.009}
\newcommand{\rseiEplugGpuCoffeeIntEight}{1.480}

\newcommand{\rseiSpeedupIntEightCpuCoffee}{1.27}
\newcommand{\rseiSpeedupGpuVsCpuCoffeeFpThirtyTwo}{1.04}

\newcommand{\rseiEraplCpuJasperFpThirtyTwo}{2.117}
\newcommand{\rseiEplugCpuJasperFpThirtyTwo}{2.938}

\newcommand{\rseiEraplCpuJasperFpSixteen}{2.187}
\newcommand{\rseiEplugCpuJasperFpSixteen}{3.014}

\newcommand{\rseiEraplCpuJasperIntEight}{1.249}
\newcommand{\rseiEplugCpuJasperIntEight}{1.650}

\newcommand{\rseiEraplGpuJasperFpThirtyTwo}{0.393}
\newcommand{\rseiEplugGpuJasperFpThirtyTwo}{0.541}

\newcommand{\rseiEraplGpuJasperFpSixteen}{0.400}
\newcommand{\rseiEplugGpuJasperFpSixteen}{0.544}

\newcommand{\rseiEraplGpuJasperIntEight}{0.497}
\newcommand{\rseiEplugGpuJasperIntEight}{0.668}

\newcommand{\rseiSpeedupIntEightCpuJasper}{1.45}

\newcommand{\rseiSpeedupGpuVsCpuJasperFpSixteen}{4.38}
\newcommand{\rseiEffGpuVsCpuJasperFpSixteen}{5.46}

\newcommand{\rseiSinconDeltaPctJasperCpuFpThirtyTwo}{+1.0}
\newcommand{\rseiSinconDeltaPctJasperCpuFpSixteen}{+0.1}
\newcommand{\rseiSinconDeltaPctJasperCpuIntEight}{+6.8}
\newcommand{\rseiNeoDriverJasper}{23.43.027642}
\newcommand{\rseiOpenClCJasper}{OpenCL C 1.2}
\newcommand{\rseiOpenVinoPinJasper}{2024.6}

\newcommand{\rseiAccMicroFpThirtyTwo}{0.5494}
\newcommand{\rseiAccMicroStdFpThirtyTwo}{0.0022}
\newcommand{\rseiAccMacroFpThirtyTwo}{0.3352}
\newcommand{\rseiAccMacroStdFpThirtyTwo}{0.0006}
\newcommand{\rseiAccDMicroFpThirtyTwo}{0.0000}
\newcommand{\rseiAccMicroFpSixteen}{0.5494}
\newcommand{\rseiAccMicroStdFpSixteen}{0.0022}
\newcommand{\rseiAccMacroFpSixteen}{0.3352}
\newcommand{\rseiAccMacroStdFpSixteen}{0.0006}
\newcommand{\rseiAccDMicroFpSixteen}{0.0000}
\newcommand{\rseiAccMicroIntEight}{0.5500}
\newcommand{\rseiAccMicroStdIntEight}{0.0022}
\newcommand{\rseiAccMacroIntEight}{0.3358}
\newcommand{\rseiAccMacroStdIntEight}{0.0007}
\newcommand{\rseiAccDMicroIntEight}{+0.0006}

\newcommand{\rseiDatasetTrain}{20000}
\newcommand{\rseiDatasetTest}{10000}
\newcommand{\rseiDatasetCalib}{512}
\newcommand{\rseiDatasetClasses}{19}

\newcommand{\rseiProtoWarmup}{20}  
\newcommand{\rseiProtoNIter}{200}  
\newcommand{\rseiProtoRepeats}{5}  
\newcommand{\rseiProtoPowerInterval}{0.5}  
\newcommand{\rseiProtoIdleSecs}{60}  

\newcommand{\rseiPatchesPerHourFloor}{71{,}600}  
\newcommand{\rseiPatchesPerHourStep}{135{,}500}  

\begin{document}
	
		\title{Quantized Edge Inference of a Self-Supervised SAR--Optical Encoder
		on Commodity x86 Hardware}
	
	\author{Christian Sanchez-Leon\orcidlink{0009-0005-0994-9946} and Dongming Chen%
		\ifpreprint
		\thanks{This work was supported by the Key Technologies Research and
			Development Program of Liaoning Province under Grant 2024JH2/102400072
			and by the Fundamental Research Funds for the Central Universities under
			Grant N25BSS062. (Corresponding author: Dongming Chen.)}%
		\else
		\thanks{Manuscript received XX; revised XX. This work was supported by the
			Key Technologies Research and Development Program of Liaoning Province
			under Grant 2024JH2/102400072 and by the Fundamental Research Funds for
			the Central Universities under Grant N25BSS062. (Corresponding author:
			Dongming Chen.)}%
		\fi
	\thanks{The authors are with the Software College, Northeastern University,
			Shenyang 110819, China (e-mail: chendm@mail.neu.edu.cn).}%
	\ifpreprint\thanks{This work has been submitted to the IEEE for possible
			publication. Copyright may be transferred without notice, after which this
			version may no longer be accessible.}\fi}
	
	\markboth{IEEE Geoscience and Remote Sensing Letters}%
		{Sanchez-Leon \MakeLowercase{\textit{et al.}}: Quantized Edge Inference of a SAR--Optical Encoder}
	
	\maketitle
	
	\begin{abstract}
		Self-supervised vision transformers are increasingly used as reusable encoders
		for Earth observation, but their inference cost on the hardware available in the
		field is rarely characterized. We deploy a pretrained SAR--optical
		masked-autoencoder encoder on three commodity x86 nodes, from a low-power CPU
		without AVX2 to an AVX2+VNNI reference, at FP32, FP16 and INT8 on the CPU and,
		where present, the integrated GPU, and report latency, energy (RAPL and wall
		plug) and linear-probe accuracy. Quantization is accuracy-neutral: INT8 changes
		micro-mAP by only \rseiAccDMicroIntEight{}, within the seed spread, so precision
		becomes a pure cost decision. INT8 lowers CPU latency on every node, by
		\rseiSpeedupIntEightCpuJasper$\times$ even on the CPU without AVX2, and CPU
		energy wherever it is measured, and the VNNI reference
		(\rseiSpeedupIntEightCpuDev$\times$) suggests a further gain beyond AVX2. The
		Pareto-optimal device shifts from the integrated GPU, which is
		\rseiSpeedupGpuVsCpuJasperFpSixteen$\times$ faster than the CPU on the weakest
		node, to the CPU as the CPU class improves.
	\end{abstract}
	
	
	\section{Introduction}
	\label{sec:intro}
	Self-supervised pretraining is now a standard way to build reusable encoders for
	Earth observation, where unlabeled Sentinel imagery is abundant but labels are
	scarce. Masked autoencoders~\cite{he2022mae} over the vision
	transformer~\cite{dosovitskiy2021vit} are a common choice, evaluated on archives
	such as BigEarthNet~\cite{sumbul2019bigearthnet} and its refined release
	reBEN~\cite{clasen2024reben}; the encoder studied here~\cite{sanchezleon2026strata}
	follows this recipe on a joint SAR--optical corpus.
	
	Most of this work is evaluated on accuracy alone, yet an encoder is only useful
	if one inference is cheap enough for the hardware actually available, often a
	small, passively cooled x86 box rather than a server GPU. The cost of that inference,
	and how it responds to post-training quantization, depends on details that are
	seldom reported: whether the CPU offers wide SIMD (AVX2) or integer
	dot-product instructions (VNNI), and whether a usable integrated GPU (iGPU) is
	present. These features vary widely across commodity hardware, so a single
	benchmark number does not tell a practitioner what to deploy.
	
	Running learned models close to the sensor is already an active topic in Earth
	observation, but most of the work targets dedicated accelerators on board.
	$\Phi$-Sat-1 flew a convolutional cloud detector on an Intel Myriad~2 vision
	processing unit~\cite{giuffrida2022phisat}, RaV{\AE}n sizes a variational
	autoencoder for change detection on resource-limited flight
	hardware~\cite{ruzicka2022raven}, and Ziaja et al.\ benchmark deep models for
	on-board use~\cite{ziaja2021benchmarking}. On the model side, post-training
	quantization of vision transformers is known to be delicate, since post-softmax,
	post-GELU and LayerNorm activations are hard to quantize, and dedicated schemes
	have been proposed for it~\cite{yuan2022ptq4vit,lin2022fqvit}. Much less is known
	about the general-purpose x86 hardware available on the ground, in small,
	passively cooled boxes, some of which still lack AVX2, and about how a
	self-supervised SAR--optical transformer behaves there when latency, energy and
	accuracy are measured together.
	
	This letter characterizes the deployment of this encoder across a gradient of
	commodity x86 classes. Our contributions are: (i) we show that quantization is accuracy-neutral for this
	SAR--optical encoder, so precision becomes a pure cost decision; (ii) we
	characterize that cost across an ISA-capability gradient and show that INT8 pays
	off even on a CPU without AVX2, with a present-day VNNI reference indicating a
	further gain beyond AVX2; and (iii) using a reproducible RAPL plus
	wall-plug protocol, we show that the Pareto-optimal device is not fixed but moves
	from the iGPU to the CPU as the CPU class improves.
	
	\section{Materials and Methods}
	\label{sec:method}
	
	\subsection{Encoder, Intermediate Representation and Task}
	We study the encoder released with~\cite{sanchezleon2026strata}, a ViT-S/16
	masked autoencoder~\cite{he2022mae,dosovitskiy2021vit} pretrained in a
	self-supervised manner on the STRATA-S12 SAR--optical corpus and archived with
	it~\cite{sanchezleon2026corpus}. The input stacks
	six co-registered optical and SAR channels and the model returns a 384-D class
	token; we deploy only the encoder, with no decoder and no masking. The encoder is
	rebuilt from timm~\cite{wightman2019timm} transformer blocks, exported to ONNX and compiled to
	OpenVINO~\cite{openvino} intermediate
	representations (IR) at FP32, FP16 and INT8. INT8 uses post-training
	quantization~\cite{jacob2018quant} with
	NNCF~\cite{kozlov2020nncf} on a real calibration subset of
	\rseiDatasetCalib{} reBEN patches, disjoint from train and test. The IRs are built
	once with the development toolchain and shipped unchanged, so a given precision
	yields identical outputs on every host: accuracy is a property of the IR and is
	measured once.
	Representation quality is assessed on reBEN~\cite{clasen2024reben} as a
	\rseiDatasetClasses-class multi-label task through a linear probe on frozen
	features (\rseiDatasetTrain{} training and \rseiDatasetTest{} test patches, three
	seeds), following the protocol of~\cite{sanchezleon2026strata} as implemented in
	its public evaluation harness~\cite{sanchezleon2026harness}; we report micro- and
	macro-mAP and the change in micro-mAP relative to FP32.
	
	\subsection{Hardware Gradient}
	We deploy on three commodity x86 nodes chosen to span a gradient of SIMD
	capability rather than of price (Table~\ref{tab:hw}): a \emph{floor} node without
	AVX2, a mainstream AVX2 \emph{step} node, and an AVX2+VNNI \emph{reference} node,
	names used throughout the letter. The first two also carry an
	Intel integrated GPU (iGPU) and act as edge targets; the reference is used for
	CPU latency only, as a present-day point of comparison rather than as an edge
	target. The OpenVINO runtime is pinned to a single release on all three nodes so
	that the CPU and iGPU paths stay comparable.
	
	\subsection{Measurement Protocol}
	For each node, device and precision the IR is compiled and run for
	$W=\rseiProtoWarmup$ discarded warm-up passes, followed by $R=\rseiProtoRepeats$
	repetitions of $N=\rseiProtoNIter$ forward passes at batch size one (single
	stream). Latency is the wall-clock time of a repetition divided by $N$, reported
	as the mean with a 95\% confidence interval (Student's~$t$) over the repetitions.
	Batch size one matches the edge use case, where tiles arrive one at a time and
	have to be processed with low delay; throughput under batching is outside the
	scope of this letter. As the network is dense and the input shape is fixed,
	latency does not depend on the pixel values. Energy is sampled on the host clock
	nominally every \rseiProtoPowerInterval~s from two synchronized sources, the
	CPU-package counter of RAPL~\cite{khan2018rapl} and the active power of a
	wall-plug meter, and is reported per inference after subtracting an idle
	baseline measured over \rseiProtoIdleSecs~s:
	\begin{equation}
		e = \frac{E_{\mathrm{window}} - \bar{P}_{\mathrm{idle}}\,\Delta t}{N},
		\label{eq:energy}
	\end{equation}
	where $E_{\mathrm{window}}$ is the energy over a repetition of duration
	$\Delta t$ and $\bar{P}_{\mathrm{idle}}$ the idle power. The CPU governor is set
	to \texttt{performance}. All parameters are defined in the released
	code\footnote{\url{https://github.com/csanchez-research/rs-edge-infer}} and
	recorded with every result (Section~\ref{sec:repro}).

\begin{table}[H]
	\centering
	\caption{Commodity x86 nodes, ordered by ISA class. The role names are used
		in the text, tables and figures.}
	\label{tab:hw}
	\footnotesize
	\setlength{\tabcolsep}{3pt}
	\begin{tabular}{@{}lllccl@{}}
		\toprule
		Role & Machine & CPU & AVX2 & VNNI & iGPU \\
		\midrule
		Floor & Moort M1 & \begin{tabular}[t]{@{}l@{}}Celeron N5105\\(Jasper Lake)\end{tabular} & No & No & UHD Gen11 \\
		Step & Lenovo M720q & \begin{tabular}[t]{@{}l@{}}i3-9100T\\(Coffee Lake)\end{tabular} & Yes & No & UHD\,630 Gen9.5 \\
		Ref. & Dev.\ laptop & \begin{tabular}[t]{@{}l@{}}i7-14700HX\\(Raptor Lake)\end{tabular} & Yes & Yes & --- \\
		\bottomrule
	\end{tabular}
\end{table}

	\section{Results}
	\label{sec:results}
	
	\subsection{Quantization is Accuracy-Neutral}
	Table~\ref{tab:accuracy} reports the linear-probe accuracy. FP16 matches FP32
	exactly (micro-mAP \rseiAccMicroFpThirtyTwo), and INT8 differs by only
	\rseiAccDMicroIntEight{} in micro-mAP, well inside the
	$\pm\rseiAccMicroStdIntEight$ seed spread, so precision is a pure cost decision.
	
	\begin{table}[H]
		\centering
		\caption{Accuracy (linear probe, 3 seeds); quantization is accuracy-neutral.}
		\label{tab:accuracy}
		\begin{tabular}{@{}lccc@{}}
			\toprule
			Precision & micro-mAP & macro-mAP & $\Delta$ micro vs FP32 \\
			\midrule
			FP32 & $\rseiAccMicroFpThirtyTwo\pm\rseiAccMicroStdFpThirtyTwo$ & $\rseiAccMacroFpThirtyTwo\pm\rseiAccMacroStdFpThirtyTwo$ & \rseiAccDMicroFpThirtyTwo \\
			FP16 & $\rseiAccMicroFpSixteen\pm\rseiAccMicroStdFpSixteen$ & $\rseiAccMacroFpSixteen\pm\rseiAccMacroStdFpSixteen$ & \rseiAccDMicroFpSixteen \\
			INT8 & $\rseiAccMicroIntEight\pm\rseiAccMicroStdIntEight$ & $\rseiAccMacroIntEight\pm\rseiAccMacroStdIntEight$ & \rseiAccDMicroIntEight \\
			\bottomrule
		\end{tabular}
	\end{table}
	
	\subsection{INT8 Lowers Cost on the CPU; FP16 Does Not}
\begin{figure}[H]
		\centering
		\includegraphics[width=0.9\columnwidth]{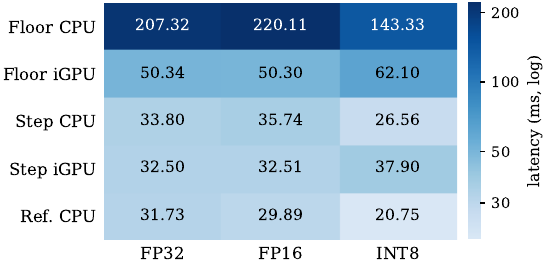}
		\caption{Per-inference latency (ms, mean over $R=\rseiProtoRepeats$
			repetitions; log color scale) by node, device and precision. On the floor node
			the iGPU is the fastest device; on the step and reference nodes INT8 on the CPU is.}
		\label{fig:heatmap}
	\end{figure}

	On the CPU, FP16 brings no benefit: it is essentially tied with FP32, and on the
	step node it is even slightly slower (\rseiLatCpuCoffeeFpSixteen{} vs
	\rseiLatCpuCoffeeFpThirtyTwo{}~ms). INT8, in contrast, is faster on every node:
	\rseiSpeedupIntEightCpuJasper$\times$ over FP32 on the floor node,
	\rseiSpeedupIntEightCpuCoffee$\times$ on the step node and
	\rseiSpeedupIntEightCpuDev$\times$ on the reference (Fig.~\ref{fig:heatmap}).
	Energy tracks latency both at the CPU package and at the wall plug
	(Table~\ref{tab:energy}), so INT8 is also the most efficient CPU precision on the
	two nodes with energy measurements.
	
	\subsection{INT8 Pays Off Without AVX2; VNNI Adds More}
	The floor node, which has neither AVX2 nor VNNI, gains
	\rseiSpeedupIntEightCpuJasper$\times$ from INT8, more than the step node
	(\rseiSpeedupIntEightCpuCoffee$\times$). This is consistent with a memory-bound
	forward pass at batch size one, where the fourfold smaller INT8 weights help even
	without wide integer instructions. Against the reference, the step node has
	almost the same FP32 latency (\rseiLatCpuCoffeeFpThirtyTwo{} vs
	\rseiLatCpuDevFpThirtyTwo{}~ms, the latter with a wider
	$\pm$\rseiLatCiCpuDevFpThirtyTwo{}~ms interval) and also carries AVX2, but its
	INT8 latency is clearly higher (\rseiLatCpuCoffeeIntEight{} vs
	\rseiLatCpuDevIntEight{}~ms). With the runtime and the IR held identical, the
	additional INT8 acceleration on the reference points to VNNI rather than to AVX2.
	
	\subsection{The Best Device Depends on the Hardware Class}
	On the floor node the iGPU is far faster and more efficient than the CPU
	(\rseiSpeedupGpuVsCpuJasperFpSixteen$\times$ faster and
	\rseiEffGpuVsCpuJasperFpSixteen$\times$ more efficient at the package at FP16; at the wall plug,
	\rseiEplugGpuJasperFpSixteen{} vs \rseiEplugCpuJasperFpSixteen{}~J per inference), so the best
	configuration there is the iGPU, at FP32 or FP16 alike. On the step node, whose CPU is much
	stronger, the same iGPU barely helps
	(\rseiSpeedupGpuVsCpuCoffeeFpThirtyTwo$\times$ at FP32), and INT8 on the CPU
	becomes the fastest configuration (\rseiLatCpuCoffeeIntEight{}~ms). Its energy
	matches FP16 on the iGPU at the package (\rseiEraplCpuCoffeeIntEight{} vs
	\rseiEraplGpuCoffeeFpSixteen{}~J) and is slightly lower at the wall plug
	(\rseiEplugCpuCoffeeIntEight{} vs \rseiEplugGpuCoffeeFpSixteen{}~J). On both nodes INT8 is the slowest precision on
	the iGPU, since these integrated GPUs run FP16 natively and gain
	nothing from integer weights. The Pareto-optimal deployment is thus not fixed:
	it shifts from the iGPU to the CPU as the CPU class improves
	(Fig.~\ref{fig:heatmap}, Table~\ref{tab:energy}).
	
	\begin{table}[H]
		\centering
		\caption{Per-inference energy (J), net of the idle baseline, for the CPU package
			(RAPL) and the whole system (wall plug). The lowest value per node and meter is
			in bold. The reference node has no energy measurement.}
		\label{tab:energy}
		\footnotesize
		\setlength{\tabcolsep}{3pt}
		\begin{tabular}{@{}llcccccc@{}}
			\toprule
			& & \multicolumn{3}{c}{RAPL (J)} & \multicolumn{3}{c}{Wall plug (J)} \\
			\cmidrule(lr){3-5}\cmidrule(l){6-8}
			Node & Device & FP32 & FP16 & INT8 & FP32 & FP16 & INT8 \\
			\midrule
			Floor & CPU  & \rseiEraplCpuJasperFpThirtyTwo & \rseiEraplCpuJasperFpSixteen & \rseiEraplCpuJasperIntEight & \rseiEplugCpuJasperFpThirtyTwo & \rseiEplugCpuJasperFpSixteen & \rseiEplugCpuJasperIntEight \\
			& iGPU & \textbf{\rseiEraplGpuJasperFpThirtyTwo} & \rseiEraplGpuJasperFpSixteen & \rseiEraplGpuJasperIntEight & \textbf{\rseiEplugGpuJasperFpThirtyTwo} & \rseiEplugGpuJasperFpSixteen & \rseiEplugGpuJasperIntEight \\
			Step  & CPU  & \rseiEraplCpuCoffeeFpThirtyTwo & \rseiEraplCpuCoffeeFpSixteen & \textbf{\rseiEraplCpuCoffeeIntEight} & \rseiEplugCpuCoffeeFpThirtyTwo & \rseiEplugCpuCoffeeFpSixteen & \textbf{\rseiEplugCpuCoffeeIntEight} \\
			& iGPU & \rseiEraplGpuCoffeeFpThirtyTwo & \rseiEraplGpuCoffeeFpSixteen & \rseiEraplGpuCoffeeIntEight & \rseiEplugGpuCoffeeFpThirtyTwo & \rseiEplugGpuCoffeeFpSixteen & \rseiEplugGpuCoffeeIntEight \\
			\bottomrule
		\end{tabular}
	\end{table}
	\subsection{Implementation and Reproducibility}
	\label{sec:repro}
	Two safeguards support reproducibility. First, every artifact along the chain
	(checkpoint $\rightarrow$ ONNX $\rightarrow$ per-precision IR) is recorded in a
	SHA-256 manifest, and a provenance gate re-checks this chain on disk before each
	run, so every result traces back to the exact inputs that produced it. Second, the
	integrated-GPU runtime is not assumed to work: we sweep OpenVINO and driver
	combinations on the floor node and pin the first release (\rseiOpenVinoPinJasper)
	whose kernels the Gen9/11 iGPU can compile, as newer releases regressed a
	normalization kernel to OpenCL~2.0 work-group functions that these devices
	(driver \rseiNeoDriverJasper, \rseiOpenClCJasper) cannot build. A version-control
	check confirms the pin does not bias the CPU numbers: FP32 and FP16 latency are
	unchanged across releases (\rseiSinconDeltaPctJasperCpuFpThirtyTwo\,\% and
	\rseiSinconDeltaPctJasperCpuFpSixteen\,\%) while INT8 shifts by
	\rseiSinconDeltaPctJasperCpuIntEight\,\%. On the reference node the same check moves FP32, FP16 and INT8
	by \rseiSinconDeltaPctDevCpuFpThirtyTwo, \rseiSinconDeltaPctDevCpuFpSixteen{} and
	\rseiSinconDeltaPctDevCpuIntEight\,\%; all nodes are therefore measured on the
	pinned release. IRs, calibration data and code are at the URL above.
	
	\section{Discussion and Conclusion}
	\label{sec:conclusion}
	Because INT8 does not cost
	accuracy, precision can be chosen on cost alone, and the right choice depends on
	the node. On a weak CPU without AVX2 the integrated GPU is the clear option, with
	FP32 and FP16 equivalent there, whereas on a mainstream AVX2 CPU the integrated GPU adds
	little and INT8 on the CPU is the fastest configuration and at least as
	efficient as the integrated GPU. INT8 is faster on the CPU of every node, and on AVX2 CPUs
	VNNI appears to add a further gain, so CPUs that expose it are the better choice
	when INT8 is planned.
	
	In practical terms, at batch size one the floor node processes about
	\rseiPatchesPerHourFloor{} patches of $224\times224$ pixels per hour on its iGPU,
	at \rseiEplugGpuJasperFpSixteen{}~J each at the wall plug, and the step node about
	\rseiPatchesPerHourStep{} per hour on its CPU in INT8, at
	\rseiEplugCpuCoffeeIntEight{}~J each.
	
	We study a single ViT-S/16 encoder, so
	the accuracy-neutral result and the magnitude of the speedups may differ for
	other architectures or sizes. The reference node is a current laptop CPU,
	included to place the edge nodes against present-day hardware rather than as an
	edge target. As it is also the only node with VNNI, the VNNI observation rests on
	two non-identical machines with a similar FP32 latency, not on a single
	controlled variable. Device temperature was not recorded, so thermal
	throttling cannot be ruled out directly, although the narrow confidence intervals
	across repetitions on the edge nodes show no drift within a run. Extending this characterization to a set of
	encoders, and to newer integrated GPUs and low-power CPUs that expose VNNI, is
	left to future work.
	
	\bibliographystyle{IEEEtran}
	\bibliography{references}
	
\end{document}